# A scanning tunneling microscope capable of electron spin resonance and pump-probe spectroscopy at mK temperature and in vector magnetic field

Werner M. J. van Weerdenburg[1,†] Manuel Steinbrecher[1,†], Niels P. E. van Mullekom[1], Jan W. Gerritsen[1], Henning von Allwörden[1], Fabian D. Natterer[2], Alexander A. Khajetoorians[1,*]

1.  *Institute for Molecules and Materials, Radboud University, 6525 AJ Nijmegen, The Netherlands*

2.  *Department of Physics, University of Zurich, CH-8057 Zurich, Switzerland*

†   The authors contributed equally to this work.

*corresponding author: a.khajetoorians@science.ru.nl

**Abstract**

In the last decade, detecting spin dynamics at the atomic scale has been enabled by combining techniques like electron spin resonance (ESR) or pump-probe spectroscopy with scanning tunneling microscopy (STM). Here, we demonstrate an ultra-high vacuum (UHV) STM operational at milliKelvin (mK) and in a vector magnetic field capable of both ESR and pump-probe spectroscopy. By implementing GHz compatible cabling, we achieve appreciable RF amplitudes at the junction while maintaining mK base temperature. We demonstrate the successful operation of our setup by utilizing two experimental ESR modes (frequency sweep and magnetic field sweep) on an individual TiH molecule on MgO/Ag(100) and extract the effective *g*-factor. We trace the ESR transitions down to MHz into an unprecedented low frequency band enabled by the mK base temperature. We also implement an all-electrical pump-probe scheme based on waveform sequencing suited for studying dynamics down to the nanoseconds range. We benchmark our system by detecting the spin relaxation time $T_1$ of individual Fe atoms on MgO/Ag(100) and note a field strength and orientation dependent relaxation time.

**I. Introduction**

Scanning tunneling microscopy/spectroscopy (STM/STS) is the preeminent method at the characterization of the geometric, electronic, and magnetic properties of individual atoms on surfaces[1-3]. The great appeal of STM is based on its spectroscopic methods, which provides exquisite insight into the local density of states (LDOS) with an energy resolution that is limited by thermal broadening. In consequence, strong efforts were dedicated over the last decades to lower the measurement temperature of STM systems below a Kelvin[4-7] and thereby to increase the energy resolution. More recently, fueled by further interest in increased energy resolution to probe the low temperature quantum phases of matter[8-10], STM systems have been built based on a dilution refrigerator design (DR) which have achieved mK operational temperatures[11-18].

Complementary to traditional STS, spin resolved STM/STS methods have been used to study the magnetic properties of e.g. individual atoms[7, 19, 20]. Nevertheless, spin-resolved STM/STS typically measures time-averaged properties due to the poor time resolution[21, 22] that is limited by the bandwidth of the transimpedance preamplifier. However, the introduction of all-electrical excitation schemes, such as pump-probe and electron spin resonance (ESR) spectroscopy methods based on spin polarized STM (SP-STM), have been developed to probe the spin dynamics of individual magnetic atoms with vastly increased time resolution. While pump-probe methods[23] provide access to the relaxation times of the magnetization ($T_1$) of individual and coupled atoms[24-26], complementary electron spin resonance yields insight into the magnetic ground state of individual impurities[27], as well as magnetic interactions[28-30] and coherent dynamics[31-33]. However, these techniques have thus far been implemented in a temperature range between 0.5 – 4.3K[27, 34-36] and were often hampered by dissipation of the microwave power which heated up the junction. Accordingly, the low-temperature range at which new physics emerges has hitherto remained out of reach. Therefore, it would be desirable to develop time-resolved methods in DR-based systems, taking advantage of both the lower temperatures compared to traditional $^3$He STM as well as the high cooling power[12], to access the lower frequency bands and to characterize the resonant dynamics at these low energy scales.

Here, based on a recent ultra-high vacuum (UHV) spin-resolved DR-STM operational at mK temperature and in vector magnetic field[18, 37], we detail the implementation and performance of both ESR and pump-probe methods at mK temperature in this system. We demonstrate ESR-STM operation in two different modes, namely frequency sweep (*f*-sweep) or magnetic field sweep (*B*-sweep). Utilizing an individual TiH molecule adsorbed on MgO/Ag(100) as a benchmark, we achieved signals down to unprecedented low radio frequency (RF) bands of only hundreds of MHz, corresponding to a lowest absolute energy scale of a 1.6 μeV[38], and with appreciable RF power to perform ESR. Observation of ESR in this frequency range demonstrates operation comparable to the expected thermal energy of our DR, enabling ESR-STM to take advantage of the mK temperature regime. Furthermore, we use the high frequency capability to apply short voltage pulses to perform pump-probe experiments and measure the field-dependent spin relaxation time $T_1$ of a single Fe atom on MgO/Ag(100).

## II. High-frequency implementation & characterization

The detailed construction of the UHV-STM system and laboratory used for RF implementation and characterization is described in reference [18]. The system houses a dilution refrigerator (Janis Research Company – JDR-50) with a base temperature of 30 mK, and is equipped with a two-axis vector magnet. The magnetic field can be up to 9 T out-of-plane ($B_\perp$) and up to 4 T in-plane ($B_\parallel$) and is 360° rotatable within one plane with a magnitude of up to 3 T. We used SPECS Nanonis RC5/SC5 for the STM control electronics and two different preamplifiers for the current detection, a Femto LCA-1K-5G in Section III or a Femto DLPCA-200 in Section II and IV. The modulated current signal was detected by a lock-in amplifier (Zurich Instruments - MF-DEV4463 in Section III and Stanford Research Systems - SR830 DSP in Sections II and IV) and we used a multifunctional I/O device for digitized data acquisition of the current and lock-in signal (National Instruments – USB-6251). We utilized a Cr bulk tip in all magnetic measurements, as previously used in our home-built systems for spin polarized STM[18, 37].

### A. High-frequency cabling

In order to accommodate GHz excitation with appreciable RF amplitude, we adapted high frequency cabling *in-situ* to establish tip and sample transmission lines. The instrumental setup is illustrated in Figure 1. From room temperature to the first stage (~4 K), we utilized stainless steel semi-rigid cables (COAX LTD - SC-086/50-SS-SS). From this stage to the mixing chamber stage, we used superconducting NbTi semi-rigid cables (COAX LTD - SC-086/50-NbTi-NbTi). From the mixing chamber stage, to the microscope, we used semi-rigid copper cables (Elspec DA 50047-821), and the last 20 cm, approximately, consisted of flexible coaxial cabling (Elspec MMK 5001). All *in-situ* connections, excluding the direct connections to the sample and tip, were made with SMP connectors rated to 40 GHz. Previously installed RF powder filters[18] and other low-pass filters were removed from both transmission lines.

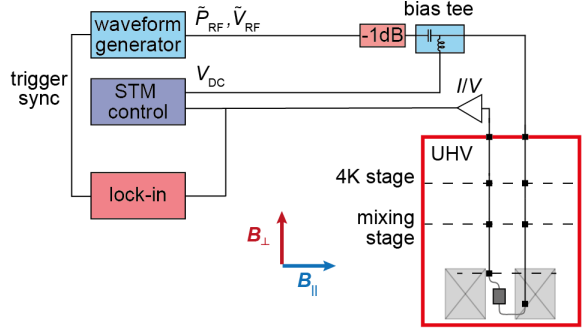

Figure 1. Sketch of the cables of the mK ESR-STM. The *in-situ* cabling has two RF lines consisting of three segments of semi-rigid cables (black), leading to tip and sample with flexible cabling (gray). The STM is enclosed by the microscope body at the center of the vector magnet.

*Ex-situ*, we used a bias tee (Tektronix PSPL5370) to combine the DC bias ($V_{DC}$) with the AC output ($\tilde{V}_{RF}$) of the RF generator (Keysight N5183B MXG Microwave Analog Signal Generator). The bias tee also served as DC block to avoid DC voltage leakage to the RF generator. An additional 1 dB attenuator (Minicircuits FW-1+) was added to the RF line to prevent charge build-up. The *ex-situ* cabling (Pasternack PE300-120; rated to 18 GHz) was connected to the UHV system via an SMA feedthrough with a grounded shield (Allectra).

### B. Characterization of the high frequency transmission

To quantify the transmission of high frequency sinusoidal signals, we used the surface state of Ag(111) as a non-linear feature in the *I-V*-curve for rectification as described in reference [39]. The high frequency signal was sent via the tip line for the subsequent measurements, unless otherwise specified. We cleaned Ag(111) with repeated cycles of Ar+ sputtering and annealing at 560°C, and subsequently studied the sample at both $T$ = 30 mK and $T$ = 7 K.

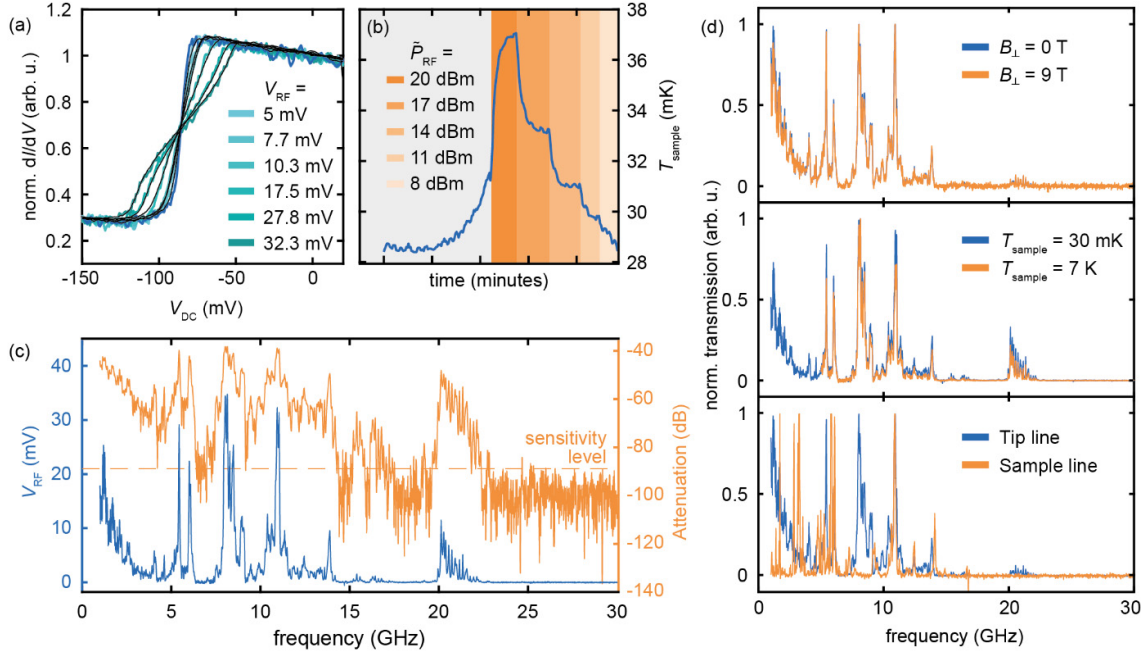

Figure 2. (a) Broadening of the Ag(111) surface state step in d$I$/d$V$ spectroscopy by application of continuous RF voltage $\tilde{V}_{RF}$ ($f$ = 10.94 GHz), added to $V_{DC}$. The broadening amplitude at the junction, $V_{RF}$ (zero-to-peak; see legend), is extracted by matching the broadened spectra with the convolution (black) of an arcsine function and the d$I$/d$V$ spectrum taken without RF irradiation (blue). (stabilized at $V_{DC}$ = -86.0 mV, $I_t$ = 1 nA; $V_{mod}$ = 1 mV, $f_{mod}$ = 4266 Hz). (b) Effect of RF irradiation on the sample temperature, detected next to the STM head, by applying continuous $\tilde{V}_{RF}$ with $f$ = 10.94 GHz at various $\tilde{P}_{RF}$ (smoothed). (c) Attenuation of RF frequencies between 1 and 30 GHz from the RF generator to the STM junction, and the resulting $V_{RF}$ for $\tilde{P}_{RF}$ = 19 dBm. The sensitivity level is defined as twice the standard deviation of the $V_{RF}$ signal between 25 and 30 GHz. (d) Transmission measurements comparing the tip line characteristics with $T_{sample}$ = 30 mK and $B_\perp$ = 0 T (blue) with the transmission with applied $B_\perp$ = 9 T (upper panel), the transmission measured with $T_{sample}$ = 7 K (middle panel) and the sample line transmission (bottom panel), all normalized to the highest peak amplitude for comparison. All transmission measurements were taken on the Ag(111) surface state step with $V_{DC}$ = -56.7 – -75.8 mV, $I_t$ = 1 nA and $f_{mod}$ = 877 - 887 Hz.

First, we identified the onset energy of the Ag surface state in STS. Next, we repeated the measurement to observe the change in step width for different values of $\tilde{V}_{RF}$ added to $V_{DC}$ (Fig. 2a). The surface state onset (step in d$I$/d$V$) was broadened by the attenuated RF signal at the junction, with amplitude $V_{RF}$. We fitted the broadened step by a convolution of an arcsine function (with effective amplitude $V_{RF}$) with the original spectrum. In Fig. 2a, we measured and fitted the broadened step width for several output powers between $\tilde{P}_{RF}$ = 1 dBm and $\tilde{P}_{RF}$ = 19 dBm ($\tilde{V}_{RF}$ = 2.818 V) at $f$ = 10.94 GHz and found $V_{RF}$ (zero-to-peak) up to 32.5 mV.

We next calibrated the transmission for a frequency range of 1 – 30 GHz, utilizing the method described in reference [39]. More specifically, we matched $V_{DC}$ to the Ag(111) step position to maximize rectification of the RF signal to the non-linearity in the $I/V$ curve. We modulated the output signal of the RF generator with a 50% duty cycle at a modulation frequency $f_{mod}$ ($f_{mod}$ = 887 Hz) well within the bandwidth of the preamplifier. This chopping led to a modulation in $I_t$, which we detected with a lock-in amplifier, using the trigger output of the RF generator as the external reference signal. The lock-in signal was calibrated against the extracted step broadening from STS measurements to find $V_{RF}(f)$. Fig. 2c shows the transmission characteristics between 1 and 30 GHz for an applied power of $\tilde{P}_{RF}$ = 19 dBm. The total attenuation A = 20*log$_{10}$($V_{RF}$ / $\tilde{V}_{RF}$) is plotted for reference, where $\tilde{V}_{RF}$ = 2.818 V corresponds to $\tilde{P}_{RF}$ = 19 dBm. Alternatively, the sharp inelastic feature of TiH on MgO was used to calibrate the transmission characteristics in measurements described later.

We found that the transmission is very robust against several variables. Firstly, we compared the RF transmission for varying magnetic field (Fig. 2d, upper panel). We found that the transmission characteristics remain unchanged in the presence of a maximum out-of-plane field ($B_\perp$ = 9 T) compared to the case where no magnetic field is applied. We also tested the cabling at two sample temperatures, namely $T$ = 7 K and $T$ = 30 mK. Similar to the previous case, there were no strong changes in transmission at these two different operational temperatures (Fig. 2d, middle panel). We note that we only measured the transmission in the $T$ = 7 K case down to 5 GHz. We also observed no significant changes in transmission for different liquid $^4$He levels.

Unlike the robust transmission characteristics against temperature and magnetic field, we found variations in the transmission characteristics if $\tilde{V}_{RF}$ was applied on the sample line compared to the tip line (Fig. 2d, bottom panel). We observed similar transmission amplitudes, but several high transmission peaks shifted to different frequencies or vanished. Likewise, we found that the low frequency transmission (~ 1 GHz) was appreciably attenuated in this case, compared to the tip line case. We note that the STM resides in a small metallic cavity, and therefore cavity modes may explain the robustness of certain high transmission bands. We utilized the tip transmission line, due to overall better transmission characteristics, especially for the frequency bands around 1 and 8 GHz.

In order to characterize the effect of RF irradiation on the base temperature, we applied a continuous RF signal with variable $\tilde{P}_{RF}$ at the frequency with the highest transmission ($f$ = 10.94 GHz), as shown in Fig. 2b. We recorded the temperature over the course of 2.5 hours for several values of $\tilde{P}_{RF}$. In Fig. 2b, the shaded regions correspond to the amplitude of $\tilde{P}_{RF}$ as well as the corresponding irradiation time. Upon irradiation of $\tilde{P}_{RF}$ = 20 dBm at $f$ = 10.94 GHz ($V_{RF}$ = 36.3 mV), the temperature raised only by $\Delta T \approx$ 8 mK and reached a steady state after 15 minutes.

### III. Benchmarking ESR-STM at mK temperature

To benchmark ESR (section III) and pump-probe spectroscopy at mK temperatures (section IV), we co-deposited Fe and Ti on two monolayers (ML) of MgO on Ag(100) (Fig. 3a), as described in refs. [27, 29]. Our sample preparation details can be found in ref. [38]. After deposition, we observed individual Fe atoms (Fig. 3b) and TiH molecules (Fig. 3c) with characteristic apparent heights (Fe: 151± 8 pm, TiH: 103 ± 8 pm). They can also be spectroscopically fingerprinted by their inelastic excitations (Fe: spin excitation at ±14.5 mV, TiH: inelastic feature at approximately ± 90 mV). We investigated both Fe and TiH adsorbed on an oxygen (top) binding site. We prepared the spin polarized tip by picking up several Fe atoms from the surface to form a few-atom cluster at the tip apex. The magnetic Cr tip orients the magnetization of the cluster, with a residual magnetic stray field $\vec{B}_{tip}$ as sketched in Fig. 3d. The magnetically stable tip enables measurements with low external magnetic fields. The base temperature for all experiments in section III and IV was 30 mK ≤ $T_{base}$ ≤ 55 mK.

#### A. ESR setup & measurement scheme

The traditional picture of the detection of the ESR signal relies on a resonance-induced change in the population distribution of the spin states[27, 40]. When the frequency matches with the energy separation between the two spin states given by the Zeeman splitting, the resonance condition ($hf = g\mu_B SB$) is satisfied and coherently driven transitions between the two spin states are induced. In the ideal case, based on a traditional two-state system where the spin resides in the ground state absent of excitation, the RF irradiation results in a 50% occupation of each state. The spin population difference can be detected with a magnetic probe, utilizing the concept of SP-STM previously utilized on individual atoms[20]. The change in spin population, due to resonant excitation, leads to a variation in the spin polarized tunneling current. The resulting ESR signal, detected as described in Section II B, will be referred to as $\Delta I_{ESR}$. We note that the spin polarization of the tunneling current is dictated by the energy dependent LDOS of both tip and sample, thus this detection depends on the given $V_{DC}$.

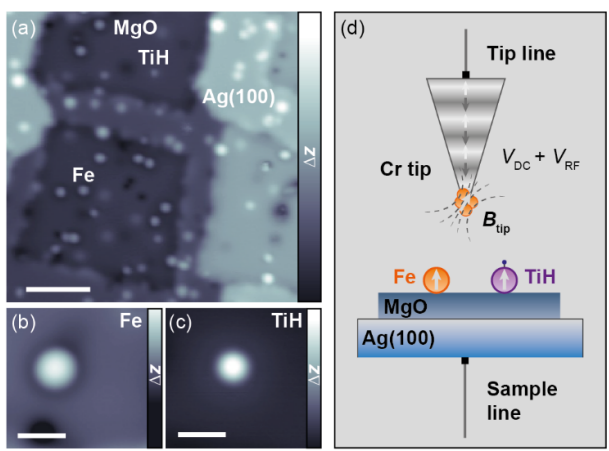

Figure 3. (a) Constant-current image of 2 ML MgO islands (scale bar = 5.0 nm, $\Delta z$ = 542 pm) on Ag(100) with individual deposited impurities. Zoom-in constant current images of (b) Fe atoms (scale bar = 1.0 nm, $\Delta z$ = 374 pm) and (c) TiH molecules (scale bar = 1.0 nm, $\Delta z$ = 125 pm) ($V_{DC}$ = 100 mV, $I_t$ = 10 pA). (d) Sketch of the sample system and the Cr bulk tip, prepared with a few-atom Fe cluster at its apex for spin polarized measurements.

In the aforementioned ideal case, the lowest detection frequency for ESR is limited by temperature. When the thermal energy exceeds the Zeeman energy, the spin can be incoherently excited leading to a thermal population of excited states[40], resulting in a decreased $\Delta I_{ESR}$[33]. Thus far, ESR-STM has been applied down to frequencies of ≈ 8 GHz[41]. The thermal energy at $T$ = 30 mK corresponds to a frequency of ≈ 0.63 GHz, thereby allowing access to much lower frequency bands compared to previous measurements, without suffering from intensity loss due to thermal effects. At the same time, this permits ESR measurements at external magnetic fields of only several mT.

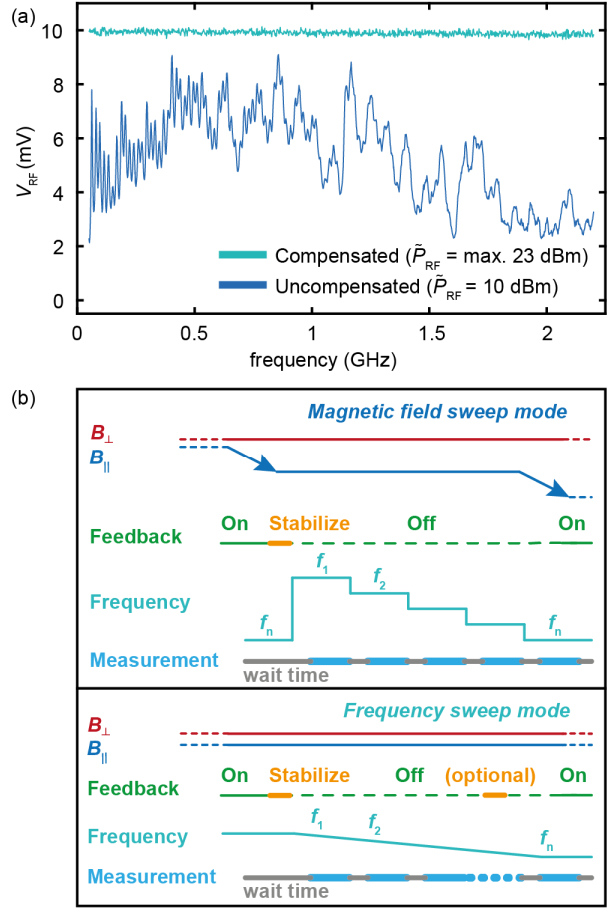

Figure 4. (a) Uncompensated $V_{RF}$ (dark blue) for frequencies between 50 MHz and 2.2 GHz ($\tilde{P}_{RF}$ = 10 dBm), detected as a rectifying signal on a spectral non-linearity on TiH ($V_{DC}$ = - 89.5 mV, $I_t$ = 200 pA, $f_{mod}$ = 877 Hz). A constant $V_{RF}$ at the junction was realized by compensating $\tilde{P}_{RF}$ for the transmission characteristics, here offset to the largest achievable constant $V_{RF}$ of 9.9 mV (light blue) in this frequency range for a maximum $\tilde{P}_{RF}$ = 23 dBm. (b) Sketch of the timing sequence for an ESR measurement in *B*-sweep mode (single iteration, upper panel) and *f*-sweep mode (lower panel).

Next, we used the *ex-situ* setup as sketched in Fig. 1 for the ESR-STM experiment. Prior to ESR detection, we recalibrated and adjusted for the non-trivial transmission function to obtain a constant $V_{RF}$ in the frequency range of interest, as described above and in ref. [39]. Fig. 4a shows an example of $V_{RF}(f)$ for frequencies between 0.05 to 2.2 GHz, measured at $\tilde{P}_{RF}$ = 10 dBm, before accounting for transfer losses. With a maximum available generator power of $\tilde{P}_{RF}$ = 23 dBm, we compensated for the transmission to maintain a maximum constant $V_{RF}$ of 9.9 mV at the tunneling junction. Power compensation was not essential for magnetic field sweeps, but calibration was required to compare the intensity of $\Delta I_{ESR}$ at different frequencies.

The ESR experiments were performed in two different measurement modes: either sweeping the frequency[27] (*f*-sweep) or magnetic field[34] (*B*-sweep). The particular algorithm for each measurement mode is detailed in each individual subfigure of Fig. 4b. Both approaches keep one parameter (total applied field *B* or *f*, respectively) constant, and the other parameter is swept in small increments (e.g., 1 MHz; 0.5 mT). When the resonance condition is satisfied, a peak in $\Delta I_{ESR}$ can be observed at the corresponding *f* or *B,* depending on the mode. We used automated software based on LabVIEW to organize communication between the required devices as well as to control the measurement sequence. We could sweep the magnetic field for both directions while remaining in tunneling contact, which reduced acquisition time. In *B*-sweep mode, we measured at multiple frequencies that had high transmission characteristics, before sweeping to the next field value. We implemented wait times after each magnetic field increment before opening the feedback loop, to ensure proper stabilization before each subsequent measurement. Since the resonance position is extremely sensitive to the tip stray field, namely the precise vertical and lateral position of the tip[27], we employed an atom tracking routine to minimize the relative changes of the tip location during the measurement. For each data point, the current and lock-in signal were recorded with an open feedback loop for 0.9 seconds and subsequently averaged. We probed frequency bands, in *f*-sweep mode, in the range of [0 - 2.2 GHz], [7.9 – 8.5 GHz] and [10.8 – 11.1 GHz] where we have sufficient transmission for constant amplitude sweeps. Measurements in *f*-sweep mode were significantly faster and were less perturbative to the temperature compared to the *B*-sweep mode.

### B.  ESR on TiH molecules on MgO on Ag(100).

We first created a magnetic tip with appreciable spin polarization by adding Fe atoms to the tip until we observe indications of spin polarization on Fe or TiH on the MgO surface[42, 43]. Below we detail the benchmarking of the ESR-STM on a TiH molecule, as reported in ref. [38]. After stabilization at the starting magnetic field, we positioned the tip at the center of the TiH molecule with $V_{DC}$ = 50 mV and $I_t$ = 2 pA, and started the measurement sequence. The recorded $\Delta I_{ESR}$ in *B*-sweep mode is plotted in Fig. 5a for a $B_{\parallel}$-sweep. The $\Delta I_{ESR}$ is expressed in fA and the offset is scaled with frequency.

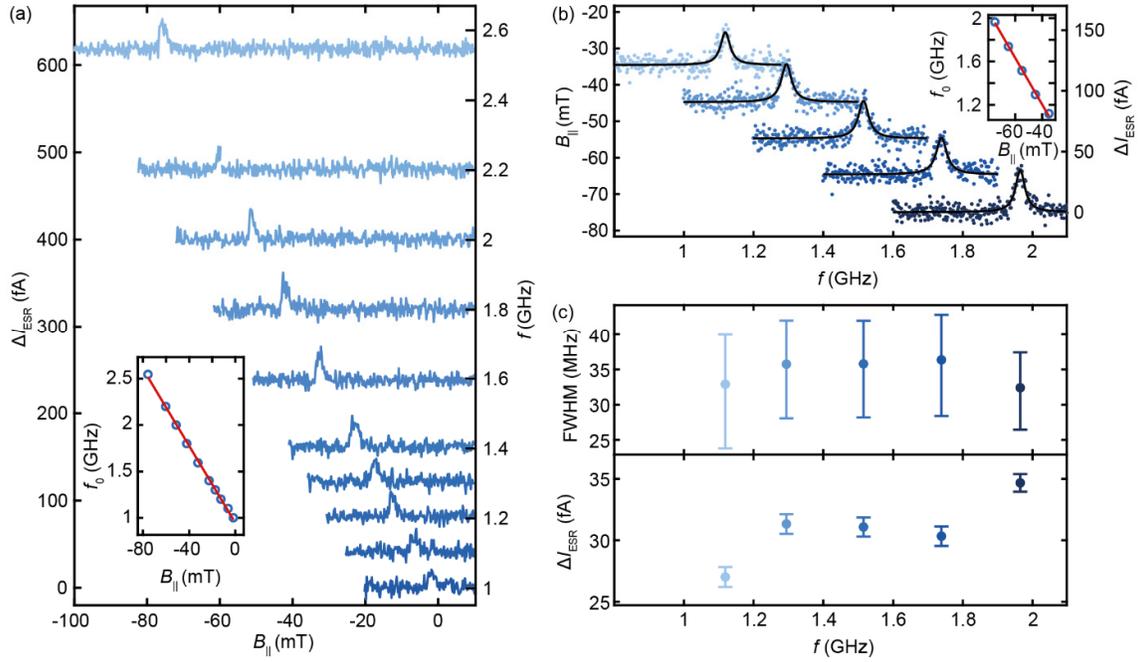

Figure 5. (a) ESR resonances of TiH/MgO measured in *B*-sweep mode by varying $B_\parallel$ ($B_\perp$ = 0 T) and applying *f* with $V_{RF}$ = 7.9 mV. Inset: resonance positions ($f_0$, $B_0$) with a linear fit to extract the *g*-factor. (b) ESR resonances on TiH/MgO measured in *f*-sweep mode by varying *f* ($V_{RF}$ = 7.5 mV) and applying constant $B_\parallel$ ($B_\perp$ = 0 T). Inset: resonance positions ($f_0$, $B_0$) with a linear fit to extract the effective *g*-factor. (c) Peak width (upper panel) and peak intensity (lower panel) extracted from the resonances in (b) by fitting a Lorentzian lineshape. The error bars represent a 95% confidence interval. ESR signals have been offset artificially. ($V_{DC}$ = 50 mV, $I_t$ = 2 pA, $f_{mod}$ = 877 Hz).

When the Zeeman splitting generated by the external magnetic field and $\vec{B}_{tip}$ matches an applied frequency we observed a strong resonance in $\Delta I_{ESR}$ of few tens of fA amplitude. We observed a linear shift as a function of frequency with a slope of (21.6 ± 0.3) GHz/T for the in-plane direction. This corresponds to an effective *g*-factor of $g_\parallel$ = 1.48 ± 0.05. The precision in *B*-sweep mode is here determined by the number of field increments, in the present case giving an accuracy of about 1 mT for each resonance position.

In Fig. 5b, we illustrate measurements made in *f*-sweep mode in the low-frequency band from *f* = 0.82 GHz up to *f* = 2.05 GHz, obtained with the same tip and molecule as used for Fig. 5a. We observed resonances in $\Delta I_{ESR}$ with similar linear Zeeman trends as in the *B*-sweep mode (for some datasets down to resonance frequencies of 382 MHz[38]). The resonances were fitted with Lorentzian lineshapes to extract their peak position, FWHM and intensity in Fig. 5c. A linear fit of the peak positions gives $g_{||}$ = 1.54 ± 0.12, and the resonances have a mean FWHM of 35 MHz and a mean intensity of 31 fA, similar to previous reports[27, 29]. Such linewidths allow for the detection of magnetic exchange, dipolar, as well as hyperfine interactions[28, 29]. However, no hyperfine splitting was observed on any of the investigated oxygen site TiH molecules. We note that the angle between the orientation of $B_{||}$ and the oxygen rows of the MgO surface is 23.8 degrees. As detailed in ref. [38], we found a statistical variation in $g$-factor values, depending on the exact tip apex, the molecule, its environment, and the systematic variation in the measurement[38], with a mean value for $g_{||}$ that agrees with literature[29]. Additionally, we observed that the tip stray field acts as offset magnetic field with values on the order of tens of mT and has a considerable influence when probing the low frequency bands.

### IV. Benchmarking pump-probe spectroscopy at mK temperature

In addition to probing ESR, we implemented an all-electrical pump-probe scheme to measure the spin relaxation times, as first described in ref. [23], at mK temperature. In such pump-probe experiments, a series of two voltage pulses with different amplitudes, the pump and probe pulse ($V_{pump}$ and $V_{probe}$), are separated by a variable delay time $\Delta t$ and periodically transmitted into the tunneling junction. First, $V_{pump}$ initializes the spin state at $t = t_0$ by overcoming the spin excitation energy ($V_{exc}$) which pumps the spin across its anisotropy barrier. Then, $V_{probe}$ is used to read out the spin state at $t = t_0 + \Delta t$ by detecting the spin polarized tunneling current while keeping $V_{probe} < V_{exc}$. By stroboscopically probing the spin state for increasing $\Delta t$, the relaxation time of the spin can be extracted from an exponential decay function.

#### A. Technical setup and characterization

To perform the pump-probe experiments, a two-channel arbitrary waveform generator (AWG) (Keysight 33622A) was added to our setup, as illustrated in Fig. 6a. A mechanical relay switch controlled whether the bias voltage generated by the STM control unit or the pulse waveform from the AWG was applied. We obtained cleaner signals by placing a variable attenuator at the output of the AWG. The attenuation was set to -30 dB, to make use of the full 14-Bit resolution of the AWG to scale down pulse amplitudes to the mV range. The DC bias and pulse waveforms were applied to the tip line. For the lock-in detection of the tunneling current, the sync input of the lock-in amplifier was connected to the trigger output of the AWG to provide an external reference signal. The *ex-situ* cabling consisted solely of standard BNC cables. The AWG itself was connected via Ethernet to the measurement computer and remote controlled via MATLAB commands.

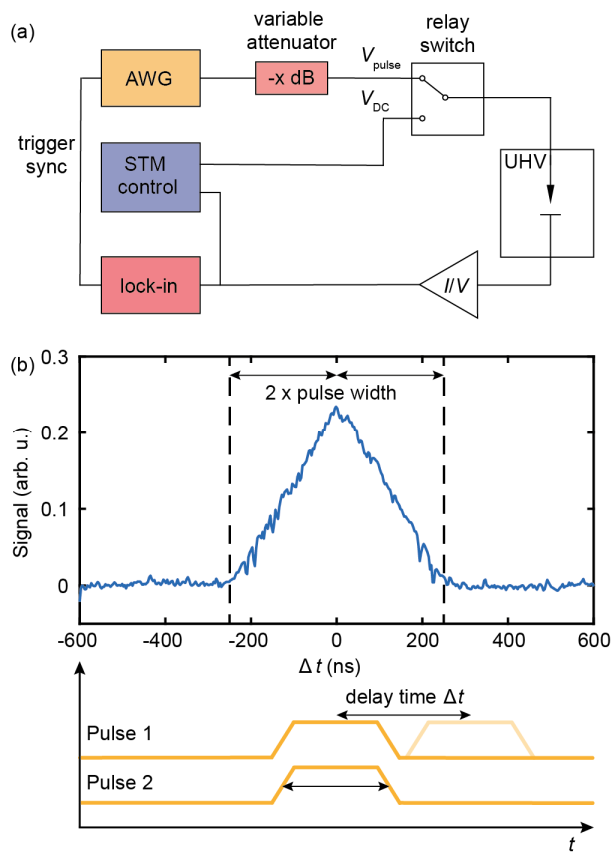

Figure 6. (a) Sketch of the pump-probe measurement set-up for the mK ESR-STM. (b) Autocorrelation measurement on Fe/MgO with two identical pulses ($|V_{pulse}|$ = 10 mV, $t_{pulse}$ = 250 ns, $r_{pulse}$ = 3.5 ns, probe span = 1200 ns), where Pulse 1 is shifted with delay time $\Delta t$ as sketched below the plot ($f_{mod}$ = 779 Hz, variable attenuator: - 30 dB).

Within this approach, we used the waveform-sequencing capability of the AWG combined with phase- and amplitude modulation of the pump and probe pulse, as described in ref. [44]. First, $V_{pump}$ with desired width $t_{pump}$, rise time $r_{pump}$ and amplitude $|V_{pump}|$ was created on channel two of the AWG. Similarly, $V_{probe}$ was prepared on channel one with width $t_{probe}$, rise time $r_{probe}$, but $|V_{probe}|$ was modulated in polarity with $f_{mod}$ well within the bandwidth of the preamplifier and slow enough to include multiple waveforms per modulation period. This resulted in two distinct cycles, A and B, that modulated $I_t$ for subsequent lock-in detection. Ultimately, both waveforms were combined internally and applied to the output of channel one. To change the time delay $\Delta t$ between $V_{pump}$ and $V_{probe}$, the position of $V_{pump}$ was shifted within the total probe span via phase modulation in a window of ±180°. The time-averaged, spin polarized tunneling current was then recorded via the synchronized lock-in amplifier for every $\Delta t$.

In contrast to ESR, which operates on continuous wave excitation at a given frequency, pump-probe spectroscopy entails transmission of more complex waveforms. In order to inspect the integrity of the waveform at the junction, we measured the autocorrelation function as previously done in refs. [23, 44]. The autocorrelation was measured by employing the pump-probe scheme across a non-linearity in the $I/V$-curve, but with two identical pulses ($t_{pump} = t_{probe}$, $r_{pump} = r_{probe}$ and $V_{pump} = V_{probe}$). The non-linearity rectifies the signal and, in the ideal case, the measured correlation shows a triangular signature of width $2 \cdot t_{pulse}$, when the identical pulses are shifted relative to each other.

The result of such an autocorrelation measurement is shown in Fig. 6b. The spin excitation of an Fe atom on MgO, as described before, was used as the non-linearity (Fig. 7a). The two pulses were created with $t_{pulse}$ = 250 ns, $r_{pulse}$ = 3.5 ns and $|V_{pulse}|$ = 10 mV. A triangular shaped waveform is identified, indicating that at time-scales on the order of 250 ns, the pulse was adequately transmitted to the junction with high fidelity. Similar autocorrelation measurements performed on Fe/Pt(111)[45] (not shown) revealed strong deviations from the triangular shape for pulse widths of ~ 20 ns, indicating the need for pulse shaping efforts to access those shorter timescales[46].

### B. Measuring spin relaxation times of an Fe atom on MgO

To benchmark pump-probe spectroscopy at mK temperatures, we used individual Fe atoms deposited on 2 ML MgO (Fig. 3b), which is an $S$ = 2 system with strong out-of-plane anisotropy[47] (see Fig. 7b). The spin relaxation time ($T_1$) between the two lowest states in the presence of a magnetic field, is on the order of microseconds[25]. With appreciable spin polarization, the ISTS spectra (Fig. 7a) were strongly modified due to significant spin pumping effects as seen in ref. [43], also indicating long-lived excited spin states.

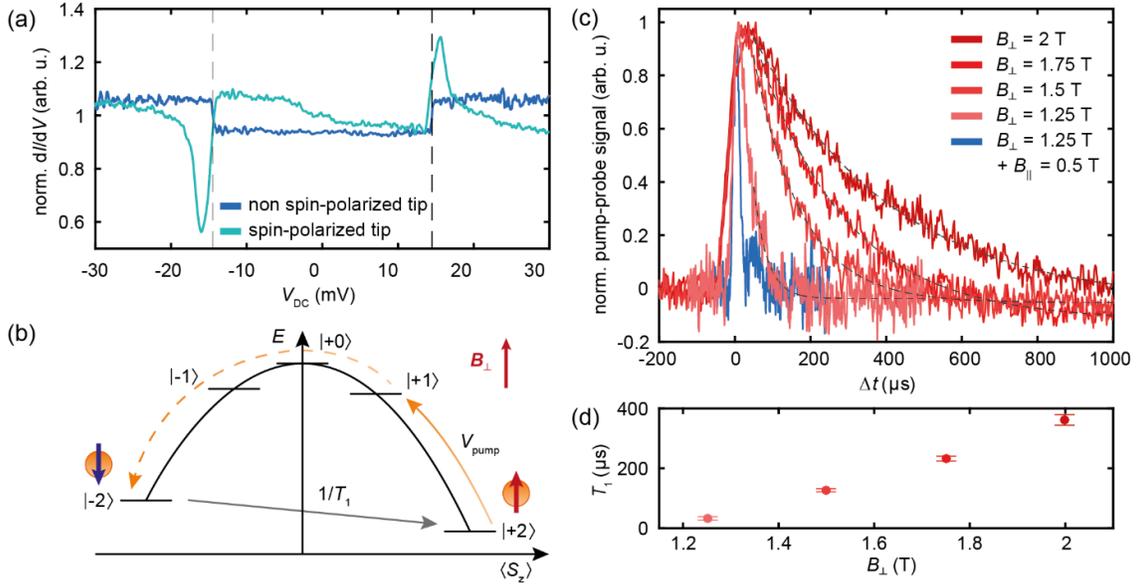

Figure 7. (a) d$I$/d$V$ spectroscopy on Fe/MgO without (dark blue) and with (light blue) a spin polarized tip. For comparison, the spectra are normalized to substrate spectra, drift corrected and smoothed for clarity. The dashed lines at ±14.8 mV indicate the spin excitation energy ($V_{exc}$). (non-SP: stabilized at $V_{DC}$ = 30 mV, $I_t$ = 500 pA, $V_{mod}$ = 150 μV, $f_{mod}$ = 809 Hz; SP: stabilized at $V_{DC}$ = 30 mV, $I_t$ = 100 pA; $V_{mod}$ = 1 mV, $f_{mod}$ = 809 Hz, $B_\perp$ = 0.145 T). (b) Sketch of the Fe spin energy levels ($S$ = 2) with an out-of-plane magnetic anisotropy axis ($z$) and lifted ground state degeneracy by Zeeman splitting ($B_\perp \neq 0$ T). The pump excitation pathway ($V_{pump} \geq V_{exc}$) and relaxation pathway are indicated by the orange and gray arrow(s) respectively. (c) Pump-probe signal as a function of probe delay time for various externally applied magnetic fields (see legend), normalized to the maximum value. The waveform sequence used for the $B_\perp$ ($B_\perp + B_\parallel$) measurement had a probe span of 2000 μs (500 μs) and included pulses with amplitudes $|V_{pump}|$ = 30 mV and $|V_{probe}|$ = 5 mV ($|V_{pump}|$ = 30 mV and $|V_{probe}|$ = 1 mV), a pulse width of $t_{pump} = t_{probe}$ = 50 μs (12.5 μs) and $r_{pump} = r_{probe}$ = 5 ns. (stabilized at $V_{DC}$ = - 5 mV , $I_t$ = 50 pA; $f_{mod}$ = 67 Hz, variable attenuator: - 30 dB). (d) Extracted relaxation time $T_1$ as a function of $B_\perp$ by fitting an exponential decay function (dashed black line in (c)).

We applied an out-of-plane magnetic field $B_\perp$ to lift the degeneracy between the two ground states $S_z = |+2\rangle$ and $|-2\rangle$. Then, $V_{pump}$ was applied to excite the spin with a certain probability from $|+2\rangle$ with subsequent $\Delta m = -1$ transitions across the anisotropy barrier into the $|-2\rangle$ state. Eventually, the spin

relaxed back into the ground state, described by a rate $1/T_1$. For larger $B_\perp$, $T_1$ increased due to the diminishing overlap between the two lowest states, which reduces the tunneling matrix elements. Measurements for different values of the applied magnetic field $B_\perp$ are shown in Fig. 7c. $B_\perp$ was varied within a range of 1.25 – 2.0 T and the $T_1$ time was probed with pulse waveforms as indicated in the figure caption. The experimental data was fitted with an exponential decay function[23, 25] (dashed black lines), excluding the data for $\Delta t \leq 50$ μs. We obtained $T_1(B_\perp)$ increasing from ≈ 35 to 360 μs with increasing $B_\perp$ (Fig. 7d), following the expectations of ref. [25]. Interestingly, the application of an additional in-plane magnetic field $B_\parallel$ reduced the lifetime drastically. This is a result of the increased mixing of the $|\pm 2\rangle$ ground states[24], which stems from the transversal anisotropy that the Fe atom exhibits. The results of our experiments on such a single Fe atom demonstrate that we have successfully applied the waveform-sequencing-based pump-probe method in our dilution refrigerator setup and demonstrate the utility of a vector magnet to control the $T_1$ time of spin systems.

**Conclusion**

In conclusion, we demonstrated GHz capabilities in a mK STM setup by upgrading the system with high frequency compatible cables. Using Fe atoms and TiH molecules on MgO, we benchmarked the performance of these upgrades. We illustrated that ESR-STM can be implemented in the mK regime, with appreciable power in a wide-range of frequency bands. We implemented two measurements modes within ESR, namely *f*-sweep and *B*-sweep, both of which provide information about the resonant excitations of individual atoms with a high degree of precision. We also demonstrated a compatible pump-probe scheme that we benchmarked on individual Fe atoms on MgO, and tune the spin relaxation time by utilizing the vector magnetic field. The pump-probe scheme was limited to rise times on the order of hundreds of ns. Future endeavors require developments such as pulse shaping at the junction, in order to reduce the effective rise time. The demonstration of GHz-STM at mK temperature provides new characterization techniques, e.g. by using photon-assisted tunneling on superconductor materials[48, 49], to probe a variety of novel phenomena in low temperatures quantum phases of matter.

**Acknowledgements**

We acknowledge funding from NWO, and the VIDI project: "Manipulating the interplay between superconductivity and chiral magnetism at the single-atom level" with project number 680-47-534. This project has received funding from the European Research Council (ERC) under the European Union's Horizon 2020 research and innovation programme (SPINAPSE: grant agreement No 818399). F.D.N. thanks the Swiss National Science Foundation for financial support under grant PP00P2_176866.

**Data availability**

The data that support the findings of this study are available from the corresponding author upon reasonable request.